\documentclass[preprint,aps,showpacs]{revtex4}

\usepackage{graphicx}

\draft 

\begin{document}


\title{Surface-directed spinodal decomposition in the pseudobinary alloy (HfO$_2$)$_x$(SiO$_2$)$_{1-x}$} 



\author{J. Liu,$^1$\email[Electronic mail:]{jliu255@uwo.ca} X. Wu,$^2$ W. N. Lennard,$^1$ D. Landheer,$^2$ and M. W. C. Dharma-Wardana$^2$}
\affiliation{
$^{1}$Department of Physics and Astronomy, University of Western Ontario, London, Ontario, Canada  N6A 3K7\\
$^{2}$Institute for Microstructural Sciences, National Research Council of Canada, Ottawa, Ontario, Canada  K1A 0R6}


\date{\today}

\begin{abstract}
Hf silicate films (HfO$_2$)$_{0.25}$(SiO$_2$)$_{0.75}$ with thicknesses in the range 4-20 nm were grown on silicon substrate by atomic layer deposition at $350\,^{\circ}\mathrm{C}$. Hf distributions in as-grown and $800\,^{\circ}\mathrm{C}$ annealed films were investigated by high resolution transmission electron microscopy (HRTEM), angle-resolved x-ray photoelectron spectroscopy (ARXPS) and medium energy ion scattering (MEIS).  HRTEM images show a layered structure in films thinner than 8 nm.  The ARXPS data also reveal a non-uniform distribution of Hf throughout the film depth.  Diffusion of SiO$_2$ to the film surface after a longer time anneal was observed by MEIS.  All these observations provide evidence for surface-directed spinodal decomposition in the pseudobinary (HfO$_2$)$_x$(SiO$_2$)$_{1-x}$ alloy system.
\end{abstract}

\pacs{68.37.Og, 64.75.St, 05.70.Fh}
\keywords{Surface-directed spinodal decomposition; Hafnium silicate films; High-$\kappa$ dielectric}

\maketitle 



\section{Introduction}

When an initially homogeneous binary mixture is rapidly quenched into an unstable state below the critical temperature, phase separation occurs via diffusion which results in a composition fluctuation throughout the system.  In the bulk mixture where the interfacial and elastic energies can be neglected, the composition fluctuation results in a random isotropic microstructure comprised of phase regions enriched in either component.\cite{Cahn1968AIME, cahn:93}  The single-phase domains then grow with time corresponding to the coarsening of the phase separated structure.\cite{bates:3258}  This phenomenon has been referred to as spinodal decomposition (SD).  In thin films where the translational and rotational symmetries are broken due to the presence of interfaces or free surfaces, spinodal decomposition may interact with wetting phenomena resulting in a very different structure at the film boundaries compared to the bulk behaviour, which has been recognized as surface-directed spinodal decomposition (SDSD).\cite{Jones:1326, Bruder:624, Krausch:3669, Puri:5359, Das:061603}  Experiments \cite{Jones:1326, Bruder:624, Krausch:3669} and simulations \cite{Puri:5359, Das:061603} have shown that in SDSD, a composition wave normal to the film surface forms at the surface due to the preferential attraction of the surface to one of the two components.  This wave then propagates into the film bulk and decays because of thermal noise.\cite{Geoghegan2003261}  Most experimental studies in SDSD have been carried out in polymer mixtures \cite{Jones:1326, Bruder:624, Krausch:3669} where the associated phase diagrams can be tailored and a small self-diffusion coefficient slows the SD dynamics.  While it is predicted that SD could occur in any two-component system whose phase diagram shows a miscibility gap, such as the ZrO$_2$-SiO$_2$, HfO$_2$-SiO$_2$, La$_2$O$_3$-SiO$_2$ or Y$_2$O$_3$-SiO$_2$ systems,\cite{kim:5094} observations of SDSD in thin solid films have not yet been reported by other groups.

As the size of complementary metal oxide semiconductor (CMOS) transistors rapidly shrinks, the thickness of the traditional gate dielectrics, i.e., SiO$_2$ and SiO$_x$N$_y$, enters into a sub-nanometer regime.\cite{packan:2079}  In this thickness range, the increase of direct tunneling current through the gate oxide raises significant power consumption and device reliability issues.  In order to reduce the gate leakage current, materials (so called high-$\kappa$) with dielectric constants larger than SiO$_2$ have been widely investigated so that a thick dielectric layer could be used as a gate insulator to improve transistor performance.\cite{wilk:5243,robertson:327} Pseudobinary alloys (ZrO$_2$)$_x$(SiO$_2$)$_{1-x}$ and especially (HfO$_2$)$_x$(SiO$_2$)$_{1-x}$, have been considered as the most promising candidates to replace SiO$_2$ and SiO$_x$N$_y$ in CMOS technology due to their thermal stability on Si and moderately high dielectric constants.\cite{wilk:484}  Amorphous thin films are suitable for CMOS transistors since grain boundaries in polycrystalline structures can introduce conducting paths.\cite{quevedo-lopez:043508}  However, phase separation has been reported in (ZrO$_2$)$_x$(SiO$_2$)$_{1-x}$ and (HfO$_2$)$_x$(SiO$_2$)$_{1-x}$ systems with x = 0.15$-$0.80 at a typical dopant activation temperature with the attendant crystallization of ZrO$_2$ or HfO$_2$.\cite{neumayer:1801, stemmer:3141}  Kim and McIntyre \cite{kim:5094} calculated the metastable extensions of the miscibility gap and spinodal for the ZrO$_2$-SiO$_2$ system based on available phase diagrams and predicted that, upon rapid thermal annealing at conventional device processing temperatures, the (ZrO$_2$)$_x$(SiO$_2$)$_{1-x}$ system with a composition in the spinodal (x=0.1$-$0.6) will separate into two phases having compositions given by such metastable extensions of the miscibility gap.  They further simulated the effect of the (ZrO$_2$)$_x$(SiO$_2$)$_{1-x}$ film/Si substrate interface on SD and predicted a composition wave normal to the substrate surface which decays when propagating into the bulk of the film.  It is expected that the (HfO$_2$)$_x$(SiO$_2$)$_{1-x}$ system would experience the same phase separation during annealing due to the similarity of the chemical properties of Hf and Zr silicates. 

Previously, we reported cross-sectional high resolution transmission electron microscopy (HRTEM) and high angle annular dark field scanning transmission electron microscopy (HAADF-STEM) images that showed SDSD in (HfO$_2$)$_{0.25}$(SiO$_2$)$_{0.75}$ films which were grown by atomic layer deposition (ALD) at $350\,^{\circ}\mathrm{C}$ and annealed at $800\,^{\circ}\mathrm{C}$.\cite{liu:041403} Here we present more experimental evidence of SDSD in these films, i.e., the line intensity profiles extracted from the cross-sectional HRTEM images, the angle-resolved x-ray photoelectron spectroscopy (ARXPS) Hf 4f peaks, medium energy ion scattering (MEIS) measurements and plan-view HRTEM images that show a layered structure for the (HfO$_2$)$_{0.25}$(SiO$_2$)$_{0.75}$ films and the growth of the composition wave during a longer time annealing.

\section{Experiments}

(HfO$_2$)$_{0.25}$(SiO$_2$)$_{0.75}$ thin films were grown on \textit{p}-type Si(100) substrates by ALD using the precursors tetrakis(diethylamido)hafnium and tris(2-methyl-2-butoxy)silanol.  The detailed growth procedure has been described previously.\cite{liu2009ecs}  ALD utilizes the self-limiting reaction mechanism between gaseous precursors and the surface species to produce a thin film one atomic layer at a time.\cite{Suntola1989261}  During a growth cycle, each precursor was introduced separately into the deposition chamber.  Alternate precursor pulses were separated by an inert gas purge step.  The precursor pulse cycles were repeated until the desired film thickness was achieved.  Prior to deposition, the substrate was heated to $500\,^{\circ}\mathrm{C}$ for 300 s in an O$_2$ environment to oxidize the H-terminated surface.  \textit{In situ} x-ray photoelectron spectroscopy (XPS) data showed that the oxidation process resulted in a $<$0.5 nm Si thermal oxide.  Films with thicknesses in the range 4-20 nm were grown while the substrates were held at $350\,^{\circ}\mathrm{C}$.  An \textit{ex situ} rapid thermal anneal (RTA) was subsequently performed in N$_2$ at $800\,^{\circ}\mathrm{C}$ for 6 s.  In order to investigate the Hf depth profiles in both the as-grown and annealed films, $\langle011\rangle$ cross-sectional transmission electron microscope (TEM) samples were prepared using standard dimpling and ion milling procedures and subsequently characterized by HRTEM and HAADF-STEM in a JEOL JEM-2100F TEM operating at 200 kV. The HRTEM and HAADF-STEM images are referred to as bright field (BF) and dark field (DF) images, respectively. The film composition variation with depth was also investigated by ARXPS using monochromatic Al \textit{K}$\alpha$ x rays.  The MEIS technique was also used to study Hf profiles in the as-grown and annealed thin films.  MEIS experiments were performed using incident 95 keV hydrogen ions wherein a double alignment (channeling/blocking) geometry ($\langle101\rangle_{in}$ and $\langle10\bar{1}\rangle_{out}$) was employed to collect the scattered ions in a two-dimensional detector.  The equipment description and experimental setup are described in detail in Kim \textit{et al}.\cite{MEISsetup} 

\section{Results and Discussion}

\begin{figure}[htbp]
\begin{center}
	\includegraphics[scale = 1]{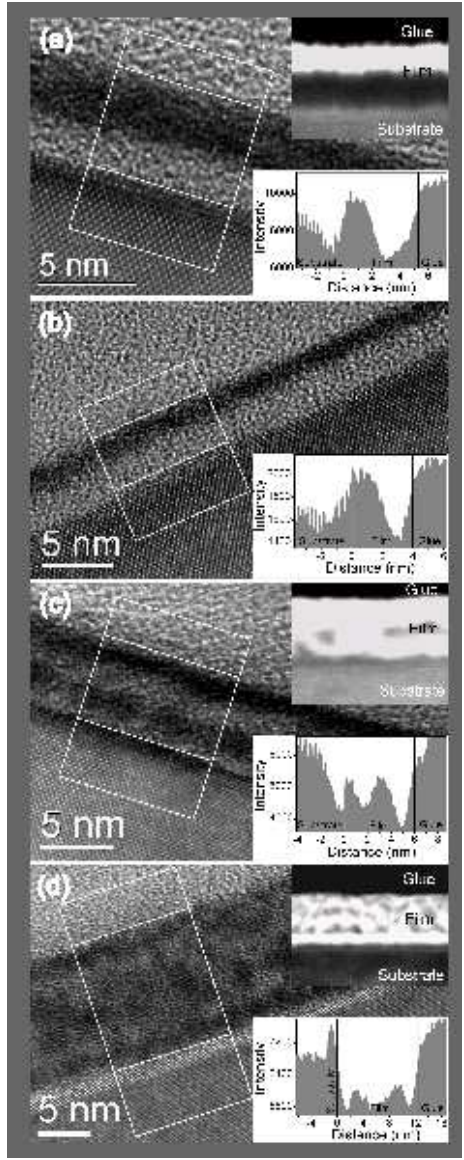}
\caption{BF images of as-grown films: (a) 5 nm;  (b) 4 nm; (c) 6.4 nm; (d) 12 nm. The upper insets show the corresponding DF images and the lower insets show the line intensity profiles integrated over the width of rectangles in the BF images.  \label{temasg}}
\end{center}
\end{figure}

Figure \ref{temasg}(a) shows the BF image of a 5 nm as-grown film. The darker area near the surface indicates a region of higher Hf concentration relative to the brighter area close to the substrate. The upper inset of Fig. \ref{temasg}(a) shows the DF image of the 5 nm film. The contrast in the DF image is reversed relative to the BF image such that the DF image is more sensitive to the atomic number of the constituent atoms. Both images confirm the Hf-rich top region and the Hf-deficient (i.e., Si-rich) bottom region of the film. The lower inset of Fig. \ref{temasg}(a) shows the line intensity profile integrated over the width of the rectangle shown in the BF image. This profile represents the Hf distribution in the direction normal to the substrate surface. A lower intensity region corresponds to a higher Hf concentration. The line intensity profile clearly shows a wave-like Hf distribution throughout the film.  The 5 nm film separates into two layers with the layer closer to the substrate Si-rich and the layer closer to the surface Hf-rich. The TEM images of thinner as-grown films (4 nm) show a similar structure, see Fig. \ref{temasg}(b).

Figure \ref{temasg}(c) shows the BF and DF images and the line intensity profile of a 6.4 nm as-grown film. The 6.4 nm as-grown film is comprised of four layers starting from the substrate:  Si-rich, Hf-rich, Si-rich and Hf-rich layers. The line intensity profile for this film [Fig. \ref{temasg}(c), lower inset] shows that the Hf-rich layer closer to the substrate has a lower Hf concentration than the Hf-rich layer closer to the film surface. The Hf concentration difference in these two layers cannot be resolved in the DF image [Fig. \ref{temasg}(c), upper inset].

As the film thickness increases to 12 nm, the layered structure can hardly be identified in the BF image [Fig. \ref{temasg}(d)]. The line intensity profile [Fig. \ref{temasg}(d), lower inset] shows that there are Hf-rich layers close to both the surface and interface. The DF image of this film [Fig. \ref{temasg}(d), upper inset] shows that there is a Si-rich layer on the substrate. This layer is followed by a Hf-rich layer. In the center of the film, Hf-rich clusters are mixed with Si-rich clusters.

The distances of the centers of the first Hf-rich layers from the substrates for the 4 nm, 5 nm, 6.4 nm and 12 nm films are 2.8, 3.5, 1.7 and 1.4 nm, respectively.  It can be concluded that as the film thickens beyond 5 nm in the film deposition process, some of the Hf atoms in the Hf-rich layer closer to the film surface diffuse towards the substrate. As already specified, the interface Si thermal oxide is $<$0.5 nm thick. Therefore, the Hf-deficient layer close to the substrate in Figs. \ref{temasg}(a) and \ref{temasg}(b) is not all thermal SiO$_2$ since its thickness is too large, i.e., significantly $>$0.5 nm, and it is a Si-rich layer of the film.

\begin{figure}[htbp]
\begin{center}
	\includegraphics[scale = 1]{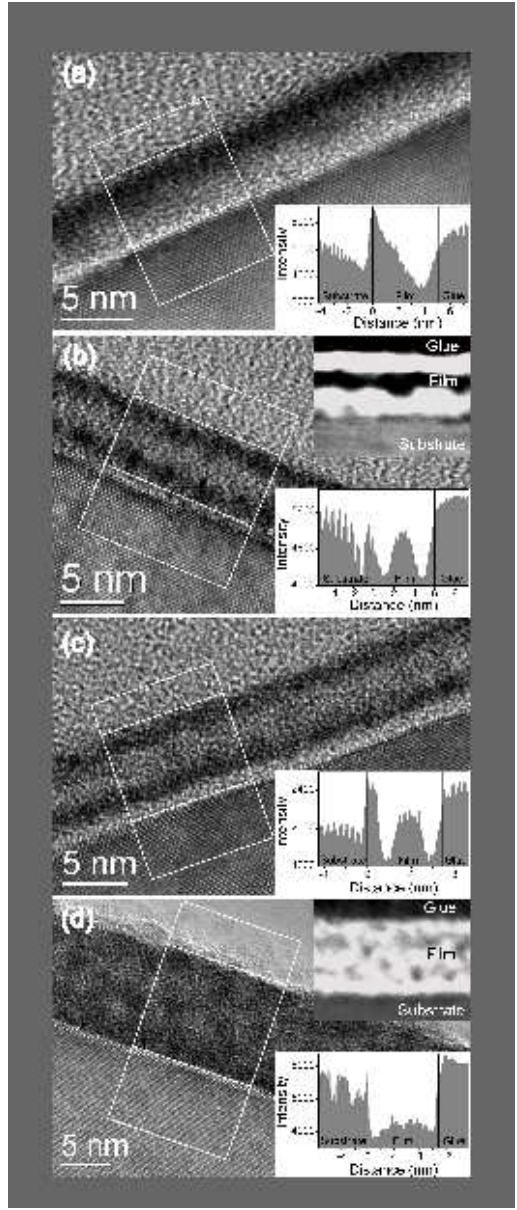}
\caption{BF images of (a) 5 nm;  (b) 6.4 nm; (c) 8 nm and (d) 12 nm films after RTA. The upper insets show the corresponding DF images and the lower insets show the line intensity profiles integrated over the width of rectangles in the BF images.  \label{temann}}
\end{center}
\end{figure}

Figure \ref{temann}(a) shows the BF image of the 5 nm film shown in Fig. \ref{temasg}(a) after RTA. The thickness of this film decreases slightly after annealing due to the densification that occurs during annealing. The film is still comprised of two layers with the layer closer to the substrate Si-rich.  The line intensity profile [Fig. \ref{temann}(a), lower inset] shows a change of the Hf distribution with respect to the profile shown in the lower inset of Fig. \ref{temasg}(a) for the as-grown film, indicating an interdiffusion of the Hf and Si atoms during the annealing process. This interdiffusion was confirmed by the binding energy (BE) shift of the Hf 4f ARXPS peaks.

\begin{figure}[htbp]
\begin{center}
\includegraphics[scale = 1]{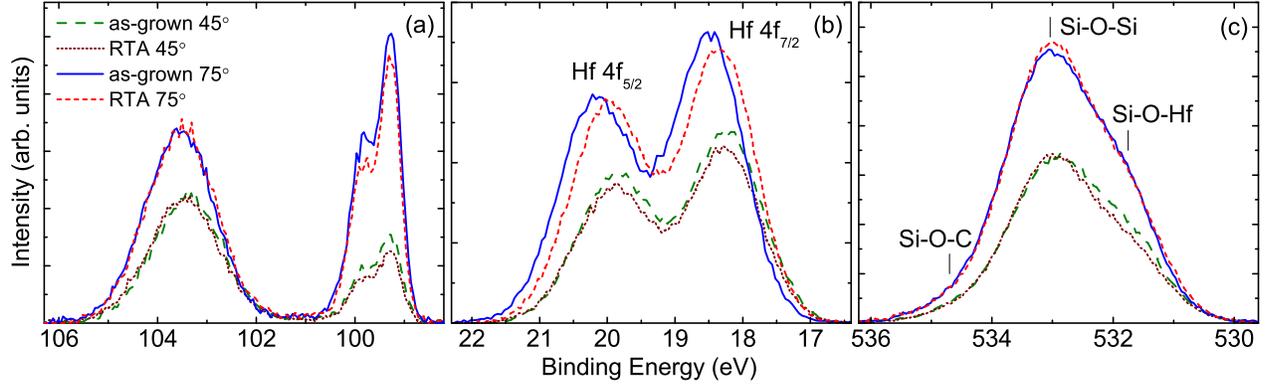}
\caption{(Color online) ARXPS peaks for the 5 nm film before and after RTA at $\theta$ = 45$^{\circ}$ and 75$^{\circ}$:  (a) Si 2p; (b) Hf 4f; (c) O 1s.  \label{arxps1355}}
\end{center}
\end{figure}

Figure \ref{arxps1355} shows the Si 2p, Hf 4f and O 1s XPS peaks at a photoelectron takeoff angle, $\theta$, of 75$^{\circ}$ and 45$^{\circ}$ for the 5 nm film before and after RTA. The Si 2p$_{3/2}$ peak from the substrate at 99.3 eV is used as a reference. The binding energies and intensities of the Si 2p peaks from the film for both photoelectron takeoff angles barely changed before and after RTA [Fig. \ref{arxps1355}(a)]. The BE of the Hf 4f$_{7/2}$ peak for the as-grown film shifts to higher energy by 0.27 eV as $\theta$ increases from 45$^{\circ}$ to 75$^{\circ}$, indicating a higher Si concentration in the film layer closer to the substrate since the ARXPS signal is more surface sensitive at lower $\theta$ and a higher Hf concentration in Hf silicate films results in a shift of BE to lower energy.\cite{ulrich:1777}  This observation is in accord with the HRTEM images of Fig. \ref{temasg}(a). The intensity of the Hf 4f peak at $\theta$ = 45$^{\circ}$ decreases after RTA, suggesting a diffusion of Hf atoms towards the substrate which results in an increase of Hf concentration in the film layer closer to the substrate, and therefore a shift (0.14 eV) of the Hf 4f peak to lower BE at $\theta$ = 75$^{\circ}$ after RTA.  The area of the O 1s peak corresponding to the Si$-$O$-$Hf bond component [Fig. \ref{arxps1355}(c), as-grown 45$^{\circ}$] also decreases after RTA, which is in agreement with the observed decrease of the Hf 4f peak intensity at $\theta$ = 45$^{\circ}$ after RTA.

Figure \ref{temann}(b) shows the BF and DF images and the line intensity profile for the 6.4 nm film after RTA. Comparing Fig. \ref{temann}(b) to Fig. \ref{temasg}(c), it can be concluded that after RTA, some of the Hf atoms in the Hf-rich layer closer to the film surface diffuse to the Hf-rich layer closer to the substrate and the distance between the first Hf-rich layer and the substrate decreases.  After RTA, the 6.4 nm film separates into three layers:  a Si-rich layer sandwiched between two Hf-rich layers. It is believed that the 0.5 nm thick interface layer in Fig. \ref{temann}(b) is Si thermal oxide formed during the substrate oxidation process before film deposition. 

Figures \ref{temann}(c) shows the BF and DF images and the corresponding line intensity profile for a film after RTA. This film is 7 nm thick after RTA $-$ 1 nm thicker than the film shown in Fig. \ref{temann}(b) $-$ and it consists of four layers starting from the substrate: Si-rich, Hf-rich, Si-rich and Hf-rich layers. 

The layered structure can hardly be identified in the BF image of the 12 nm film after RTA [Fig. \ref{temann}(d)]. The line intensity profile of this film [Fig. \ref{temann}(d), lower inset] shows that the Hf concentration near the surface and interface is slightly higher than that in the film. The DF image of this film [Fig. \ref{temann}(d), upper inset] shows that there is a Si-rich layer close to the substrate, which is followed by a Hf-rich layer. Above these two layers, Hf-rich domains are mixed with Si-rich domains. 

The above shown TEM images strongly suggest that the structure of these (HfO$_2$)$_{0.25}$(SiO$_2$)$_{0.75}$ films is caused by SDSD. A composition wave normal to the substrate surface is observed. If the composition wavelength, $\lambda_C$, is defined as the distance between the centers of two successive Hf-rich layers, then $\lambda_C$ measured from the TEM images [Figs. \ref{temann}(b) and \ref{temann}(c)] is $\sim$4 nm.  If the film is thinner than 8 nm (i.e., 2$\lambda_C$), a layered structure is observed via TEM throughout the entire film:  the surface layer is Hf rich and the layer closest to the interface Si oxide can be Si rich or Hf rich depending on the film thickness.  If the film thickness is $\lambda_C$ or 2$\lambda_C$, the layer closest to the interface is Si rich.  If the film thickness is 1.5$\lambda_C$, the layer closest to the interface is Hf rich. This result is rather surprising because the (HfO$_2$)$_{0.25}$(SiO$_2$)$_{0.75}$ films were grown on a very thin layer of Si thermal oxide. The Si-rich component in these films should have a lower interface energy with the interfacial Si thermal oxide layer. Therefore, it was to be expected that the film layer closest to the interface should always be a Si-rich layer. As the film thickens to $>$2$\lambda_C$, a Si-rich layer followed by a Hf-rich layer is observed at the film/Si interface [Fig. \ref{temann}(d)] and there is no continuous Hf-rich layer in the center of the film, indicating a tendency for decay of the composition wave.

To further study the effect of film thickness on the film structure, dilute HF (0.4$\%$) was used to etch back the 6.4 and 12 nm films. Before etching, the (HfO$_2$)$_{0.25}$(SiO$_2$)$_{0.75}$ films were subjected to the usual RTA process to reduce the etching rate \cite{chen:G483} and to achieve the layered structure shown in Fig. \ref{temann}(b) for the 6.4 nm film.  After etching, the RTA step was repeated. Spectroscopic ellipsometry was used to monitor the etching rate, (i.e., the film thickness was measured after every 4 s etch). For the 6.4 nm film, XPS was also used to measure film thickness \cite{lu:2764} before and after etching. 

\begin{figure}[htbp]
\begin{center}
\includegraphics[scale = 1]{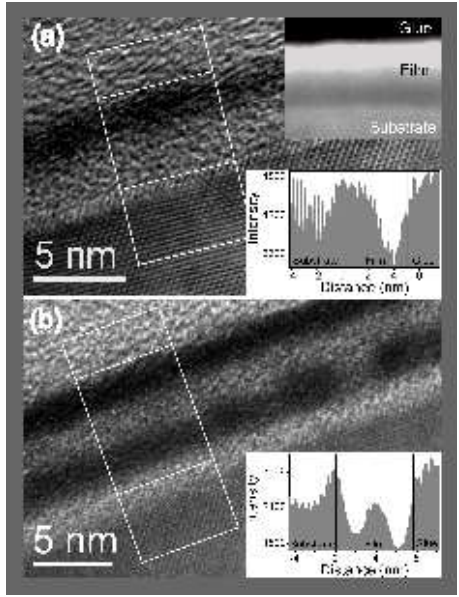}
\caption{BF images of (a) 6.4 nm, and (b) 12 nm films after RTA, HF etch and RTA. The upper inset in (a) shows the corresponding DF image and the lower insets in (a) and (b) show the line intensity profiles integrated over the width of rectangles in the BF images.  \label{temetch}}
\end{center}
\end{figure}

Figure \ref{temetch}(a) shows the BF and DF images and the line intensity profile for the 6.4 nm film after RTA, HF etch and RTA. A layer of 2 nm film was removed after the HF etch, i.e., the top Hf-rich layer in Fig. \ref{temann}(b) was removed.  The Hf-rich layer closest to the substrate in Fig. \ref{temann}(b) has diffused to the top of the film after the HF etch and following the second RTA, resulting in a structure similar to that in Fig. \ref{temann}(a), which corresponds to a 5 nm film (annealed) on a Si substrate.  The thickness of the film after RTA shown in Fig. \ref{temetch}(a) is 1 nm greater than expected after HF etching, which can likely be explained by oxidation of the Si substrate during the annealing process via water molecules absorbed in the film during the HF etching process.  

Figure \ref{temetch}(b) shows the BF image and the corresponding line intensity profile for the 12 nm film after RTA, HF etch (removing $\sim$6 nm) and RTA. The Si-rich layer closest to the substrate is probably thermal SiO$_2$ formed after HF etching and the second RTA. The line intensity profile shown in the lower inset of Fig. \ref{temetch}(b) reveals a wave-like distribution of Hf atoms in the direction perpendicular to the substrate surface. The predominantly layered structure, which was not observed for the 12 nm film [Figs. \ref{temasg}(d) and \ref{temann}(d)], appears when the film thickness is reduced to a value in the region of $\lesssim$2$\lambda_C$.

\begin{figure}[htbp]
\begin{center}
\includegraphics[scale = 1]{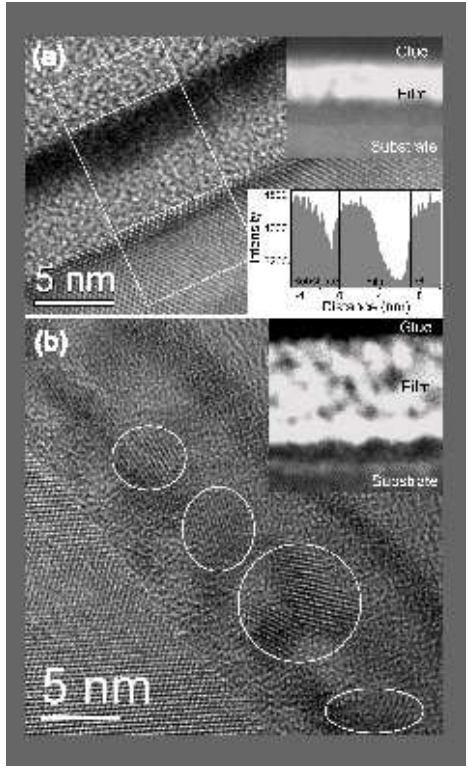}
\caption{BF images of (a) 6.4 nm, and (b) 12 nm films after a 600 s anneal at $800\,^{\circ}\mathrm{C}$ in N$_2$. The upper insets show the corresponding DF images and the lower inset in (a) shows the line intensity profile integrated over the width of rectangle in the BF image. Some of the crystalline HfO$_2$ regions are encircled.
\label{temann10}}
\end{center}
\end{figure}

According to the theory and simulation of SDSD, the single-phase domain, and therefore the composition wavelength in thin films, will increase as the annealing time increases, which corresponds to a coarsening of the phase separated structures.\cite{Puri:5359, kim:5094} In order to search for this phenomenon in Hf silicate films, longer time anneals were performed for the 6.4 nm and 12 nm films. Figure \ref{temann10}(a) shows the BF and DF images and the line intensity profile for the 6.4 nm film after a 600 s anneal at $800\,^{\circ}\mathrm{C}$ in N$_2$. Only one Hf-rich layer and one Si-rich layer are observed indicating a growth of single-phase regions during a longer time anneal. The thickness of this film has increased by $\sim$1 nm compared to the thickness after RTA [Fig. \ref{temann}(b)] as a result of diffusion of the O$_2$ impurity in N$_2$ to the interface where oxidation of the substrate occurs.\cite{maria:3476} 

\begin{figure}[htbp]
\begin{center}
	\includegraphics[scale = 1]{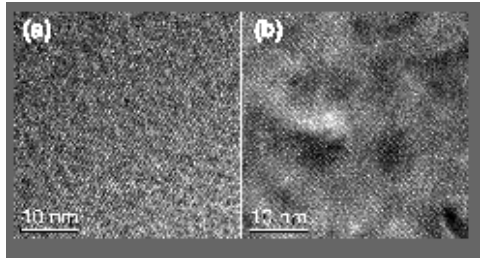}
\caption{Plan-view BF images of the 6.4 nm film after (a) RTA and (b) 600 s anneal at $800\,^{\circ}\mathrm{C}$ in N$_2$.  \label{templanview}}
\end{center}
\end{figure}

The plan-view BF images presented in Figs. \ref{templanview}(a) and \ref{templanview}(b) show the 6.4 nm film after RTA and 600 s anneal at $800\,^{\circ}\mathrm{C}$ in N$_2$, respectively. These images also indicate a growth of either the Hf-rich or Si-rich domains (i.e., either the dark or the bright areas) in the plane of the film during a longer time anneal.

\begin{figure}[htbp]
\begin{center}
\includegraphics[scale = 1]{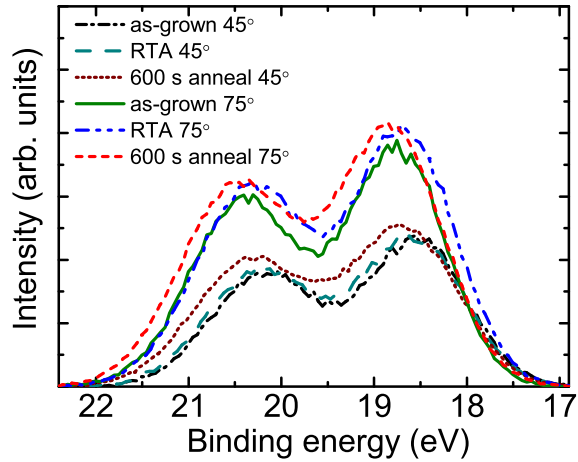}
\caption{(Color online) ARXPS Hf 4f peaks for the 6.4 nm as-grown film, and for the same film after RTA and 600 s anneal at $800\,^{\circ}\mathrm{C}$ in N$_2$.  \label{arxps1371}}
\end{center}
\end{figure}

Figure \ref{arxps1371} shows the Hf 4f ARXPS peaks (at $\theta$ = 45$^{\circ}$ and 75$^{\circ}$) for the 6.4 nm as-grown film, and for the same film after RTA and 600 s anneal at $800\,^{\circ}\mathrm{C}$ in N$_2$.  For the as-grown and RTA films from the surface to 4 nm depth, the 6.4 nm film [Figs. \ref{temasg}(c) and \ref{temann}(b)] has a similar structure as the 5 nm film [Figs. \ref{temasg}(a) and \ref{temann}(a)]:  i.e., a Hf-rich layer near the surface followed by a Si-rich layer.  The number of photoelectrons that escape from the film decreases exponentially with depth.  The contribution to the XPS peak from the Hf-rich layer closest to the substrate [Figs. \ref{temasg}(c) and \ref{temann}(b)] is negligible compared with the contribution from the layers above.  It is therefore not surprising that Hf 4f peaks from the 6.4 nm film have similar shifts as those peaks from the 5 nm film when $\theta$ changes from 45$^{\circ}$ to 75$^{\circ}$ before and after RTA.  For the as-grown film, the Hf 4f$_{7/2}$ peak shifts to higher energy by 0.23 eV as $\theta$ increases from 45$^{\circ}$ to 75$^{\circ}$.  At $\theta$ = 75$^{\circ}$, the Hf 4f$_{7/2}$ peak shifts 0.08 eV to lower energy after RTA compared with the as-grown film.  After a 600 s anneal, the Hf 4f$_{7/2}$ peaks shift to a higher BE for both $\theta$ = 45$^{\circ}$ and 75$^{\circ}$, indicating that the Hf-rich phase in the film after a longer time anneal has a lower Hf concentration than that in the as-grown film or in the film after RTA. However, the Hf 4f$_{7/2}$ peak after 600 s anneal shifts to higher BE by 0.15 eV as $\theta$ increases from 45$^{\circ}$ to 75$^{\circ}$, still indicating a higher Hf concentration in the film layer closer to the surface, which is in agreement with the TEM image [Fig. \ref{temann10}(a)]. The intensities of these Hf peaks cannot be compared directly because they are not aligned to a reference.

Figure \ref{temann10}(b) shows the BF and DF images for the 12 nm films after a 600 s anneal at $800\,^{\circ}\mathrm{C}$ in N$_2$. Comparing to the film thickness after RTA [Fig. \ref{temann}(d)], it can be concluded that a $\sim$2 nm thick layer of thermal SiO$_2$, i.e., almost all the Si-rich layer close to the substrate in Fig. \ref{temann10}(b), has grown during the 600 s anneal. HfO$_2$ crystallites were observed in this film indicating that nucleation and growth of HfO$_2$ occurred after spinodal decomposition during the annealing process. The DF image of this film [Fig. \ref{temann10}(b), upper inset] is similar to Fig. \ref{temann}(d): there is a Si-rich layer on the substrate with a $\sim$2 nm Hf-rich layer on top; above these two layers, Hf-rich clusters are mixed with Si-rich clusters. HfO$_2$ crystallites are not observed for films thinner than 12 nm even after an anneal at $800\,^{\circ}\mathrm{C}$ in N$_2$ for 600 s, which is not surprising since the onset of crystallization of HfO$_2$ in Hf silicate films depends on both film composition \cite{stemmer:3593} and thickness.\cite{pant:032901}

\begin{figure}[htbp]
\begin{center}
\includegraphics[scale = 1]{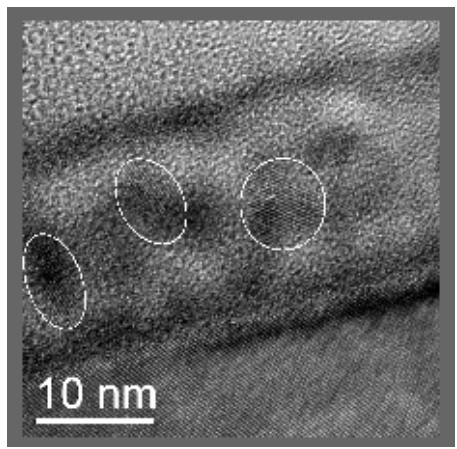}
\caption{BF images of a 20 nm as-grown films. Some of the crystalline HfO$_2$ regions are encircled. \label{tem1376}}
\end{center}
\end{figure}

In the nucleation and growth mechanism, for a nucleus to be stable with respect to further growth, it must reach a critical size. Smaller nuclei are unstable and may dissolve because of surface energy effects and their large surface to volume ratio.\cite{Lysaght2008399} Therefore the nucleation and growth process is suppressed in thinner films and a higher temperature is needed for crystallization to occur.\cite{pant:032901} In contrast, HfO$_2$ crystallites with dimension of 5$-$8 nm were observed in a 20 nm as-grown film (Fig. \ref{tem1376}), which suggests that the nucleation and growth mechanism follows the spinodal decomposition during the film deposition process at $350\,^{\circ}\mathrm{C}$. 

From those cross section TEM images, e.g., Figs. \ref{temasg}, \ref{temann}, \ref{temetch} and \ref{temann10}, it seems that the film layer closest to the surface is always a Hf-rich layer. However, if a Si-rich layer is on the surface, it is difficult to resolve this layer from the glue that is used to prepare the TEM sample. In order to examine the film composition as a function of film depth, Hf/(Hf+Si) values were extracted from ARXPS measurements and the results are presented in Table \ref{ARXPS} for various $\theta$ values.  The Hf concentration, which is proportional to the Hf/(Hf+Si) ratio, decreases towards the surface for all films (therefore implying an increase in the Si concentration).  Comparing the Hf concentration in the as-grown and annealed films at $\theta=10^{\circ}$, more SiO$_2$ is observed to diffuse to the film surface during annealing which suggests a wetting of the film surface by the SiO$_2$-rich component.  Previously, simulation results showed that the Si-rich phase contains $>$98 mol$\%$ SiO$_2$ in the (ZrO$_2$)$_x$(SiO$_2$)$_{1-x}$ system after phase separation during $900\,^{\circ}\mathrm{C}$ anneals.\cite{kim:5094}  Assuming that the Si-rich phase in the (HfO$_2$)$_{0.25}$(SiO$_2$)$_{0.75}$ films is pure SiO$_2$, then the thickness of the surface layer of the films can be estimated.  The escape depth ($\lambda_e$) of Si 2p photoelectrons excited by Al $K\alpha$ x rays in SiO$_2$ is 3 nm,\cite{lu:2764} which should be a reasonable estimate as well for Hf silicate films.  The sampling depth then varies from 9 nm (3$\lambda_e$) at $\theta=90^{\circ}$ to 1.6 nm at $\theta=10^{\circ}$.  Assuming that the film composition determined at $\theta=75^{\circ}$ corresponds to the average composition, then the thickness of the surface SiO$_2$ layer is estimated to be $\sim$0.4$-$0.7 nm,  i.e., equivalent to 1$-$2 layers of SiO$_2$.

\begin{table}[htbp]
\caption{Hf concentration extracted from ARXPS peak intensities. $\theta$ is the photoelectron takeoff angle.\label{ARXPS}}
\begin{ruledtabular}
\begin{tabular}{c c c c c}
 & & \multicolumn{3}{c}{Hf/(Hf+Si)} \\
 \cline{3-5}
 Film thickness & & \multicolumn{3}{c}{$\theta$} \\
 (nm) & Film description & 75$^\circ$ & 45$^\circ$ & 10$^\circ$ \\
\hline
5 & As-grown & 0.262 & 0.261 & 0.249 \\
5 & RTA & 0.260 & 0.249 & 0.225 \\
6.4 & As-grown & 0.242 & 0.246 & 0.228 \\
6.4 & RTA & 0.225 & 0.236 & 0.198 \\
6.4 & 600 s anneal & 0.220 & 0.223 & 0.189 \\
12 & 600 s anneal & 0.239 & 0.235 & 0.178 \\
12 & RTA, HF etch, RTA & 0.222 & 0.212 & 0.140 \\

\end{tabular}
\end{ruledtabular}
\end{table}

\begin{figure}[htbp]
\begin{center}
\includegraphics[scale=1]{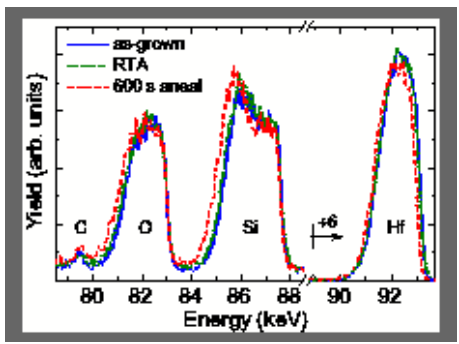}
\caption{(Color online) MEIS spectra of the 6.4 nm film:  as-grown, after RTA and after 600 s anneal at $800\,^{\circ}\mathrm{C}$ in N$_2$.  \label{meis1371}}
\end{center}
\end{figure}

MEIS measurements were also performed to study the film composition and the Hf depth profile for the 6.4 nm film.  Figure \ref{meis1371} shows the MEIS spectra for the 6.4 nm as-grown film and for the same film after RTA and a 600 s anneal at $800\,^{\circ}\mathrm{C}$ in N$_2$.  The MEIS spectra are aligned to the O edge.  As can be seen, the Si edges of the three spectra are also aligned.  The Hf edge shifts slightly to lower energy after RTA, indicating a diffusion of the HfO$_2$ component to the substrate or a diffusion of SiO$_2$ to the film surface.  After a 600 s anneal, the Hf edge shifts to lower energy by 120 eV. If the density of the surface SiO$_2$ layer is assumed to be 2.2 g/cm$^3$, then this energy shift corresponds to a diffusion of 0.35 nm SiO$_2$ to the film surface during the 600 s anneal process.  This result is in good agreement with the estimate of the surface SiO$_2$ layer thicknesses from ARXPS data and confirms that the film surface is wetted by the SiO$_2$ layer after anneal.  Compared to the as-grown film, the Si and O peaks are obviously wider after a 600 s anneal, which is due to the oxidation of the substrate by O$_2$ impurities in N$_2$, as mentioned earlier.  The slight increase of the O and Si areas after RTA probably results from the same process.  It should be pointed out that the MEIS technique measures absolute areal density (i.e., the product $\rho$t, $\rho$ is the film density and t is the film thickness) for the atomic constituents of a film.  Therefore, MEIS cannot determine the film thickness without a knowledge (or assumption) of the film density.  The MEIS data are unable to resolve the layered structures shown in Figs. \ref{temasg}(c) and \ref{temann}(b), since the ion beam extent in any direction (0.1$-$1 mm) in MEIS experiments is much larger than the composition wavelength observed for these Hf silicate films.

\begin{figure}[htbp]
\begin{center}
\includegraphics[scale=1]{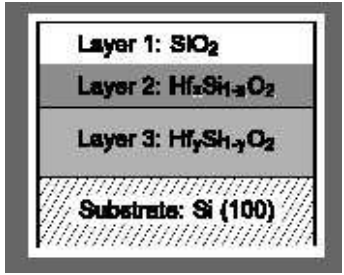}
\caption{Schematic showing the layered structure for the 5 nm film.  \label{structure}}
\end{center}
\end{figure}

The film compositions (Table \ref{ARXPS}) calculated from the ARXPS peak intensities show that the 5 nm as-grown film has the same Hf concentration at $\theta$ = 45$^{\circ}$ and 75$^{\circ}$.  However, Fig. \ref{arxps1355}(b) shows that the Hf 4f peaks of this film shift to higher energy as $\theta$ increases from 45$^{\circ}$ to 75$^{\circ}$, indicating a lower Hf concentration in the film layer closer to the substrate.  This apparent discrepancy can be explained using a simplified schematic for the Hf concentration gradient across the film depth as shown in Fig. \ref{structure}.  Layer 1 is SiO$_2$ and Layer 2 (Hf$_x$Si$_{1-x}$O$_2$) has a higher Hf concentration than Layer 3 (Hf$_y$Si$_{1-y}$O$_2$), i.e., $x > y$.  The average composition of Layers 1 and 2 is the same as the composition of Layer 3.  When the XPS measurements are taken at $\theta$ = 45$^{\circ}$, most of the signal intensity comes from Layers 1 and 2.  As $\theta$ increases to 75$^{\circ}$, the contribution of photoelectrons excited in Layer 3 increases significantly; therefore the Hf 4f peak will shift to higher energy since Layer 3 has a lower Hf concentration than Layer 2.  However, the film composition calculated from the XPS peak intensities will not change since Layer 3 has the average composition of Layer 1 and 2.  The layered structure shown in Fig. \ref{structure} agrees with the HRTEM image shown in Fig. \ref{temasg}(a).

Using grazing-incidence small angle x-ray scattering (GISAXS), Stemmer \textit{et al.} \cite{stemmer:3141} observed interference peaks in the horizontal cuts of their two-dimensional GISAXS intensity distribution, which correspond to a Hf concentration fluctuation in the plane of the film with $\lambda_C$-values of 5 nm in the 4 nm (HfO$_2$)$_x$(SiO$_2$)$_{1-x}$ films ($x$ = 0.4) after $1000\,^{\circ}\mathrm{C}$ annealing.  This observation was confirmed by their plan-view TEM image, which showed interconnected Hf-rich and Si-rich regions.  However, these authors did not try to interpret vertical cuts of their GISAXS data but still concluded that their observations were inconsistent with SDSD.  In contrast, the TEM images reported here always show a layered structure for post-annealed films of thickness $<$8 nm, which is consistent with SDSD.

A crucial observation revealed from the present study is the effect of restricted dimensionality on spinodal decomposition of HfO$_2$-SiO$_2$ systems. In effect, when the region available for phase separation is a 2-dimensional film in the $x-y$ plane with a restricted thickness $w$ in the $z$ direction, the $w$ parameter controls the ensuing structure. This can be understood by looking at the free-energy of the system which consists of species $i=1,2,\cdots$ with crystal structures $i\nu, i\mu, \cdots$. The free energy consists of a configurational energy $E(c_{1\nu},c_{2\mu},\cdots)$ based on the possible crystal structures $c_{1\nu},c_{2\mu},\cdots$ and an entropy component $S(c_{1\nu},c_{2\mu},\cdots)$. Thus the total free energy can be written as
$$F(c_{1\nu},c_{2\mu},\cdots)=E(c_{1\nu},c_{2\mu},\cdots)-TS(c_{1\nu},c_{2\mu},\cdots),$$
where $T$ is the temperature. The free energy analysis gives the equilibrium state and becomes strictly valid if the spinodal structures are the ground state of the system. If such structures result from a kinetic process (i.e., if they are metastable structures), the free energies must be replaced by energies corresponding to quasi-stationary states. However, the present analysis is adequate to understand the observed effects.

In a thick film containing two species (e.g., SiO$_2$ and HfO$_2$), the energetically favourable total nanostructure is made up of the bulk crystal structures which we denote by $c_{1b}$ and $c_{2b}$. Here, if species 1 is HfO$_2$, then $c_{1b}$ (i.e., the bulk HfO$_2$ structure of large grains) is the monoclinic structure.\cite{cheynet:054101} If the second material is Si the bulk structure is cubic, while in the case of SiO$_2$ we may take it to be the crystalobite structure or some modification thereof, as needed for surface films.\cite{PhysRevB.65.165339} We need to consider the situation where both Si atoms and Hf atoms may compete for bonding with O atoms. If thin films are formed on a substrate, then we have to deal with possible spinodals based on the bulk structures and the crystal structures manifested by vanishingly thin films. Let us denote the crystal structures found in vanishingly thin films by $c_{1f}$ and $c_{2f}$. The thinnest HfO$_2$ films tend to form orthorhombic structures,\cite{cheynet:054101} while the structure of very thin SiO$_2$ layers on Si substrates tend to be quasi-crystalline or amorphous modifications of the crystalobite structures.\cite{PhysRevB.65.165339}

When the thickness $w$ of the film is extremely small, e.g., less than a nanometer, bonds involving only a single binary species are possible in the $x-y$ plane but the $z$ direction thickness $w$ would only allow a monolayer or two of a single (binary) species such as HfO$_2$. Our experiments show the formation of a Hf-rich surface layer whose structure can be interpreted starting from an orthorhombic disposition of the Hf-O bonds but taking account of their termination at the surface. When the thickness $w$ of the deposited layer is increased, inner layers face competition, forcing them towards the bulk crystal structure $c_{1b}$, which in this case is monoclinic. A similar competitive process involving the crystal structures of species 2, i.e., SiO$_2$, also occurs. Bonds may also be formed between the common element (oxygen) with Hf or Si. However, there is not enough volume in a thin film to achieve the full configurational entropy of each species by full phase separation. Thus a compromise is achieved by SDSD. This compromise itself depends on the film thickness, partly because the limiting structures $c_{ib}$ and $c_{if}$ are different in each species. Other factors which may come into play are effects associated with crystal growth kinetics (i.e., non-equilibrium effects). However, such effects are, in our view, unimportant in this problem as the effects can be explained via a model which assumes local thermodynamic equilibrium.

The validity of the above picture is evident on a qualitative basis. A more detailed quantitative evaluation is not considered here since simulation of SiO$_2$/HfO$_2$ structures using first principles density functional methods\cite{PhysRevB.65.165339} or tight-binding methods\cite{tit:387} is beyond the scope of the present study.

\section{Conclusions}

Surface-directed spinodal decomposition in Hf silicate films during the film growth and annealing process can be envisioned from the above observations.  In the as-grown film, there is always a very thin layer of SiO$_2$ ($\sim$monolayer of SiO$_2$) at the film surface. Beneath this layer is a Hf-rich layer.  When the film thickness is $<$5 nm, Hf atoms diffuse towards the film surface as the film thickness increases, resulting in a Hf-rich layer closer to the film surface and a Si-rich layer closer to the substrate.  As the film thickness increases to values $>$5 nm, some of the Hf atoms in the Hf-rich layer closer to the film surface diffuse towards the substrate. If the thickness is 1.5$\lambda_C$, the film separates into three layers after RTA: a Si-rich layer sandwiched between two Hf-rich layers.  If the film thickness is $\sim$2$\lambda_C$, the film has a four-layer structure starting from the substrate surface:  Si-rich, Hf-rich, Si-rich and Hf-rich.  As the thickness increases to $>$2$\lambda_C$, the film loses its layered structure in the center of the film.  After a longer time anneal, the layered structure coarsens and the composition wave grows.  Crystallization of HfO$_2$ was observed in a 12 nm film after 600 s anneal at $800\,^{\circ}\mathrm{C}$ and in a 20 nm as-grown film, indicating that nucleation and growth of HfO$_2$ follows spinodal decomposition during the annealing or film deposition process.

The configurations in as-grown and annealed (HfO$_2$)$_{0.25}$(SiO$_2$)$_{0.75}$ films are qualitatively in agreement with the theory of SDSD, i.e., composition waves were observed normal to the film surface. The observation that the composition of the film layer in contact with the substrate can be affected by film thickness has never been predicted by any SDSD simulation. Presumably, theoretical studies of SDSD could be modified to accommodate these experimental results. At this time, it is difficult to study SDSD in (HfO$_2$)$_{0.25}$(SiO$_2$)$_{0.75}$ films to determine whether the composition wave obeys the growth law, i.e., $\lambda_C \sim$ t$^{1/3}$,\cite{Krausch:3669} because: (i) TEM is a qualitative technique concerning atomic composition;  (ii) SD in oxide systems is not easily controlled since the process occurs during the film deposition process;  (iii) an alternative kinetic process, viz., nucleation and growth, can impede SD during the late stages of phase separation; and (iv) any O$_2$ impurity in the annealing N$_2$ ambient may also diffuse through the film and oxidize the substrate. Although GISAXS is a powerful tool to characterize composition fluctuations in thin films,\cite{ISI:A1989CK48400003,stemmer:3141} caution has to be taken when using this technique to study SDSD in Hf silicate films since HfO$_2$ crystallizes at a relatively low temperature in thick films, as seen in the 20 nm as-grown film.  While SDSD in the (HfO$_2$)$_{0.25}$(SiO$_2$)$_{0.75}$ thin films has been confirmed, the composition range for which (HfO$_2$)$_x$(SiO$_2$)$_{1-x}$ films experience SD and the resultant compositions of the phase-separated domains are still open questions.  

The present observation of SDSD in (HfO$_2$)$_x$(SiO$_2$)$_{1-x}$ films may present significant device performance and reliability challenges for high-$\kappa$ gate dielectric applications of pseudobinary alloy systems such as ZrO$_2$-SiO$_2$, Y$_2$O$_3$-SiO$_2$, La$_2$O$_3$-SiO$_2$, etc.,\cite{kim:5094} and have effects on thin film applications for any two-component system whose phase diagram shows a miscibility gap. The ALD growth mechanism for two-component films could also be influenced by SDSD if the film surface is preferentially attracted to one of the two components. 


%
%

%


\begin{acknowledgments}
The authors are grateful for the technical assistance of L. Lebrun, G. Parent and S. Moisa at NRCC and J. Hendriks at UWO.  Financial assistance to WNL was provided by NSERC (Canada).
\end{acknowledgments}


\end{document}